\documentclass[12pt,preprint]{aastex}
\makeatletter
\renewcommand{\fnum@figure}{Figure~\thefigure}
\makeatother
\usepackage{epsfig}
\usepackage{natbib}
\usepackage{graphicx}
\usepackage{slashbox}
\usepackage{multirow}
\usepackage{lscape}
\usepackage{mathrsfs,amssymb}
\usepackage{amsmath}
\usepackage{cite}
\usepackage{float}
\usepackage[section]{placeins}
\usepackage{booktabs}
\let\captionbox\relax
\usepackage{caption}
\usepackage{enumitem}
\newcommand       \Angstrom     {\,{\rm \AA}}

\newcommand       \cm           {\,{\rm cm}}

\newcommand       \erg          {\,{\rm erg}}

\newcommand       \km          {\,{\rm km}}
\newcommand       \s            {\,{\rm s}}
\newcommand       \Hz          {\,{\rm Hz}}

\newcommand       \nH           {n_{\rm H}}
\newcommand       \NH           {N_{\rm H}}
\newcommand       \rmH         {{\rm H}}
\newcommand       \simlt        {\lesssim}
\newcommand       \simgt        {\gtrsim}

\newcommand       \mum          {\,{\rm \mu m}}
\newcommand       \ppm          {\,{\rm ppm}}
\newcommand       \Teff         {T_{\rm eff}}

\newcommand       \simali       {\sim\,}
\newcommand       \magni        {\,{\rm mag}}
\newcommand       \Fobs         {F_\lambda^{\rm obs}}
\newcommand       \Fintrinsic   {F_\lambda^{\rm int}}
\newcommand       \Frk          {F_\nu^{\rm RK}}
\newcommand       \CTOH      {\left[{\rm C/H}\right]_{\rm nano}}
\newcommand       \Ftarget    {F_{\lambda,\,{\rm targ}}}
\newcommand       \Fcomp    {F_{\lambda,\,{\rm comp}}}
\newcommand       \dcomp    {d_{\rm comp}}
\newcommand       \dtarget   {d_{\rm targ}}
\newcommand       \Accm     {A_{\lambda,\,{\rm CCM}}}
\newcommand       \constant    {{\rm const}}
\countdef\decade=200
\advance\decade by \year
\countdef\hours=201
\advance\hours by \time
\divide\hours by 60
\countdef\mins=202
\advance\mins by \hours
\multiply\mins by 60
\multiply\hours by 100
\countdef\miltime=203
\advance\miltime by \hours
\advance\miltime by \time
\advance\miltime by -\mins

\shorttitle{The Extinction Curve toward HD\,93222}
\title{
\vspace*{-2.0em}
\vspace*{1.0em}
On the Interstellar Extinction Curve
toward HD\,93222, A Sightline with
an Exceedingly Narrow 2175$\Angstrom$ Extinction Bump
}
\author{Qian~Wang\altaffilmark{1,2},
            Aigen Li\altaffilmark{2},
            and X.J.~Yang\altaffilmark{1,2}}
\altaffiltext{1}{Department of Physics,
                      Xiangtan University,
                      411105 Xiangtan, Hunan Province, China;
                      {\sf xjyang@xtu.edu.cn}}
\altaffiltext{2} {Department of Physics and Astronomy,
                  University of Missouri,
                  Columbia, MO 65211, USA
                  {\sf lia@missouri.edu}}
\begin{document}

\begin{abstract}
The 2175$\Angstrom$ extinction bump,
the most prominent spectral feature
superimposed on the interstellar extinction curve,
is widely seen in the interstellar medium (ISM)
of the Milky Way and external galaxies, both near and far.
While its central wavelength is remarkably stable and
independent with environment,
its width shows considerable variation
and environmental dependence.
Here we examine the extinction curve
for the line of sight toward HD\,93222,
a young star located in the Carina nebula.
It is found that the 2175$\Angstrom$ bump
is extremely sharp,  which is among the narrowest
ever found in the Milky Way and external galaxies.
We model the derived extinction curve and find
that, to explain the extinction characteristics
of HD\,93222, in addition to the conventional silicate
and graphite dust mixture, an additional population
of nano-sized graphitic grains is required.
\end{abstract}
\keywords {dust, extinction --- ISM: lines and bands --- ISM: molecules}

\section{INTRODUCTION\label{sec:intro}}
First detected six decades ago by Stecher (1965),
the 2175$\Angstrom$ extinction bump,
spanning roughly the wavelength range
between 1700 and 2700$\Angstrom$,
is the most prominent spectral feature
superimposed on the interstellar extinction curve.
It is now widely seen in the Milky Way
and nearby galaxies including M31,
the Large Magellanic Cloud, and several regions
in the Small Magellanic Cloud
(e.g., see Fitzpatrick 1986,
Bianchi et al.\ 1996,
Gordon \& Clayton 1998,
Misselt et al.\ 1999,
Clayton et al.\ 2000,
Gordon et al.\ 2003,
Valencic et al.\ 2004,
Fitzpatrick \& Massa 2007,
Ma\'iz-Apell\'aniz \& Rubio 2012,
Dong et al.\ 2014,
Clayton et al.\ 2015,
Decleir et al.\ 2019,
Wang et al.\ 2022).
It has also been seen in more distant galaxies,
including gravitational lens galaxies and
damped Ly$\alpha$ absorbers at redshifts
$z$\,$\simlt$\,1 (e.g., see Motta et al.\ 2002),
dusty intervening MgII systems
at $z$\,$\simlt$\,2 toward quasars
(see Wang et al.\ 2004; Srianand et al.\ 2008;
Zhou et al.\ 2010; Jiang et al.\ 2011;
Ma et al.\ 2015, 2017),
star-forming galaxies at 1\,$<$\,$z$\,$<$\,3
(e.g., see Noll et al.\ 2007, 2009;
Conroy et al.\ 2010;
Kriek et al.\ 2013;
Battisti et al.\ 2020;
Shivaei et al.\ 2022),
and gamma-ray burst host galaxies
at various redshifts
(El{\'{\i}}asd{\'o}ttir et al.\ 2009;
Liang \& Li 2009, 2010;
Prochaska et al.\ 2009;
Zafar et al.\ 2011, 2012).
In the era of the {\it James Webb Space Telescope}
(JWST), the 2175$\Angstrom$ extinction bump
has recently also been seen in the cosmic dawn
at $z$\,$\simgt$\,6 (see Witstok et al.\ 2023;
Markov et al.\ 2023, 2025a, 2025b;
Ormerod et al.\ 2025;
Fisher et al.\ 2025).

The most striking characteristics of
the 2175$\Angstrom$ extinction bump
are the invariant central wavelength
and variable bandwidth.
Its peak position at 2175$\Angstrom$
(4.6$\mum^{-1}$) is remarkably constant
and only varies  by $\pm$0.46\%
(2$\sigma$; see Wang et al.\ 2023).
In contrast, its full width half maximum
(FWHM) varies by $\pm$12\% (2$\sigma$)
around 469$\Angstrom$
($\approx$1$\mum^{-1}$;
see Fitzpatrick \& Massa 1986,
Valencic et al.\ 2004).

In this work, we explore the 2175$\Angstrom$
extinction bump for the line of sight toward
HD\,93222, a young star located in the Carina nebula,
which is among the sharpest ever found
in the Milky Way and external galaxies.
This paper is structured as follows.
We first derive in \S\ref{sec:extcurv}
the extinction curve of HD\,93222
from the far ultraviolet (UV) to the near infrared (IR).
We model the derived extinction curve
in \S\ref{sec:discussion} in terms of
the standard silicate–graphite interstellar dust model
aided by an additional nano carbon dust population.
The results are summarized in \S\ref{sec:summary}.

\section{The Extinction Curve\label{sec:extcurv}}
HD\,93222 is a young star of a spectral type of O7\,IIIf,
located in the Carina nebula, a large and complex region
of bright and dark nebulosity in the constellation Carina.
This nebula lies within the Carina–Sagittarius arm of
the Milky Way and is one of the largest star-forming
regions of the Milky Way, situated $\simali$2,600\,pc
from Earth (Markova et al.\ 2018).

We first determine the UV extinction
in the wavelength range of
$3.3 < \lambda^{-1} < 8.7\mum^{-1}$
by comparing the dust-obscured spectrum
of HD\,93222 obtained with
the {\it International Ultraviolet Explorer} (IUE)
with the intrinsic, unreddened spectra of
stars of the same spectral type as HD\,93222.
The IUE spectra were taken from the {\it Mikulski
Archive for Space Telescopes}\footnote{%
  {\sf \url{https://archive.stsci.edu/iue/}}
  },
including data from three cameras:
the Short-Wavelength Prime (SWP) Camera
covering $\simali$1150–-1978$\Angstrom$,
and the two long-wavelength cameras,
i.e., the Long-Wavelength Prime (LWP)
and the Long-Wavelength Redundant (LWR)
covering $\simali$1850–-3347$\Angstrom$.
The dust-free intrinsic stellar spectrum is
approximated by the stellar atmospheric
model spectrum of Castelli \& Kurucz (2004).\footnote{%
  {\sf http://kurucz.harvard.edu/grids/gridxxxodfnew}
  }
These stellar atmosphere models are characterized
by four parameters: the stellar effective temperature $\Teff$,
gravity $\log\,g$, metallicity ${\rm [M/H]}$,
and microturbulence $\xi$.
Taking the spectral type to be O7\,IIIf
(which determines $\Teff$),
we explore a range of stellar model parameters
for HD\,93222 by varying metallicity and gravity,
and select the combination that best fits
the observed IUE spectrum.
The stellar parameters of HD\,93222,
including $\Teff$, $\log\,g$, ${\rm [M/H]}$,
distance ($d$), luminosity ($L_\bigstar$),
and stellar radius ($R_\bigstar$)
are listed in Table~\ref{tab:stellarpara}.

The extinction at wavelength $\lambda$
is denoted by $A_{\lambda}$.
The observed IUE spectrum ($\Fobs$)
and the intrinsic, dust-free stellar spectrum
($\Fintrinsic$) are related through dust extinction
\begin{equation}\label{eq:fit}
\Fobs = \Fintrinsic
\exp \left(-\frac{A_{\lambda}}{A_{V}} \frac{A_{V}}{1.086}\right) ~.
\end{equation}

Let $\Frk$ be the Kurucz model atmospheric
flux of Castelli \& Kurucz (2004) at stellar surface
(in unit of $\erg\s^{-1}\cm^{-2}\Hz^{-1}$).
At an Earth–stellar distance of $d$,
the dust-free intrinsic flux
(in unit of $\erg\s^{-1}\cm^{-3}$) would be
\begin{equation}\label{eq:Fintrinsic}
\Fintrinsic = \Frk \times \frac{c}{\lambda^{2}}
\times \left(\frac{R_\bigstar}{d}\right)^2 ~,
\end{equation}
where $c$ is the speed of light.

Following Fitzpatrick \& Massa
(1988; hereafter FM88), we represent
the wavelength-dependence of extinction
by an analytical formula
consisting of three parts,
\begin{equation}\label{eq:Alambda2AV}
\frac{A_{\lambda}}{A_{V}} = c_{1} + c_{2} x
+ c_{3} D\left(x; \gamma, x_{0}\right) + c_{4} F(x) ~,
\end{equation}
where $x\equiv \lambda^{-1}$ is
the inverse wavelength (or wavenumber);
$c_{1} + c_{2} x$ is the linear background;
$D\left(x;\gamma, x_{0}\right)$, a Drude function
of width $\gamma$ (in unit of $\mum^{-1}$)
peaking at $x_0$ (also in unit of $\mum^{-1}$)
which characterizes the 2175$\Angstrom$
extinction bump, is defined as
\begin{equation}\label{eq:Drude}
D\left(\mathrm{x} ; \gamma, x_{0}\right)
=\frac{x^{2}}{\left(x^{2}-x_{0}^{2}\right)^{2}+x^{2} \gamma^{2}} ~;
\end{equation}
and $F(x)$ is the far-UV nonlinear rise
at $x>5.9\mum^{-1}$ as described by
\begin{equation}\label{eq:FarUV}
  F(x)=
  \begin{cases}
    0.5392(x-5.9)^{2}+0.05644(x-5.9)^{3}
    & x \geq 5.9\,\mu\mathrm{m}^{-1}  ~,\\
    0 & \mathrm{x}<5.9\,\mu\mathrm{m}^{-1} ~.
  \end{cases}
\end{equation}

\begin{figure}[htp]
\centering
\vspace{-1mm}
\includegraphics[width=\textwidth]{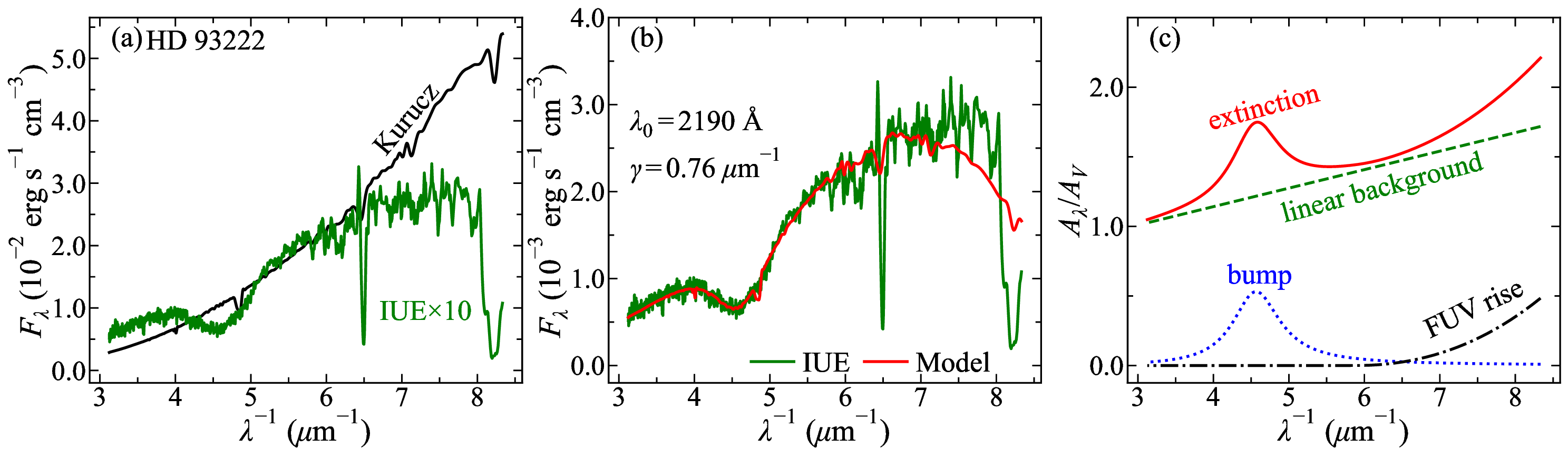}
\caption{\label{fig:HD93222}\footnotesize
Left panel (a): Comparison of the observed,
dust-obscured IUE spectrum
($\Fobs$; green solid line) of HD\,93222
with the ``intrinsic'', extinction-free spectrum
represented by the Kurucz atmospheric model
($\Fintrinsic$; black solid line).
Note that, to facilitate comparison,
the IUE spectrum is multiplied by a factor of 10.
Middle panel (b): Comparison of the observed,
dust-obscured IUE spectrum
($\Fobs$; green solid line)
with the best-fit model spectrum
($F_\lambda^{\text{int}} \exp \{-A_\lambda / 1.086\}$;
red solid line).
Right panel (c): The derived extinction curve
expressed as $A_{\lambda}$/$A_V$
for the line of sight toward HD\,93222.
The red line is the FM88 parametrization
at $\lambda^{-1} > 3.3\,\mum^{-1}$,
which is the sum of a linear ``background'' (green line),
a Drude bump of width $\gamma$
and central position of
$x_0\equiv\lambda_0^{-1}$ (blue line),
and a nonlinear far-UV rise (black dash-dotted line)
at $\lambda^{-1} > 5.9\,\mum^{-1}$.
}
\vspace{-1mm}
\end{figure}

We make use of the Levenberg–Marquardt algorithm
to fit the observed spectrum (see eq.\,\ref{eq:fit})
by minimizing ${\chi }^{2}$.
The best-fitting parameters $x_0$, $\gamma$,
$c_1$, $c_2$, $c_3$, and $c_4$
are listed in Table~\ref{tab:Fitpara}.
Figure~\ref{fig:HD93222} shows
the observed IUE spectrum ($\Fobs$)
in comparison with the Kurucz model-based,
extinction-free ``intrinsic'' spectrum ($\Fintrinsic$),
as well as the best-fitting, dust-obscured Kurucz model
spectrum ($F_\lambda^{\text{int}} \exp \{-A_\lambda / 1.086\}$).
Also shown is the derived extinction curve
and its three components, i.e., the linear background,
the 2175$\Angstrom$ bump, and the far-UV rise.
We note that this approach actually determines
$c_ j\times A_V$ (where $j$\,=\,1, 2, 3, 4),
i.e., $c_j$ and $A_V$ are degenerated.
To break this degeneracy, we adopt
$A_V=R_V\times E(B-V)\approx1.71\magni$,
where $E(B-V)\approx0.35\magni$ is
the reddening (Fitzpatrick et al.\ 2019)
and $R_V\equiv A_V/E(B-V)\approx4.76$
is the optical total-to-selective extinction ratio
(Valencic et al.\ 2004).
We will discuss the choice of $R_V$
in \S\ref{sec:environment}.

We also need to construct the extinction curve
at $\lambda^{-1} < 3.3\mum^{-1}$.
To this end, we employ the parametrization
of Cardelli, Clayton, \& Mathis (1989; hereafter CCM)
who found that most of the Galactic interstellar extinction
curves can be parametrized by a single parameter $R_V$.
For $1.1 < \lambda^{-1} < 3.3\mum^{-1}$,
we derive $A_{\lambda}/A_V$ from the CCM
parameterization for $R_V=4.76$.
However, a discontinuity often arises
between the FM88 parameterization
at $\lambda^{-1} > 3.3\mum^{-1}$
and the CCM parameterization
at $\lambda^{-1} < 3.3\mum^{-1}$.
To ensure a smooth transition and
maintain consistency with the observed
extinction-to-gas ratio $A_V/\NH$,
we apply a scaling factor to the FM88 curve,
allowing it to seamlessly match the CCM curve.

\begin{figure*}[htp]
\vspace{-1mm}
\begin{center}
\includegraphics[width=10cm,angle=0]{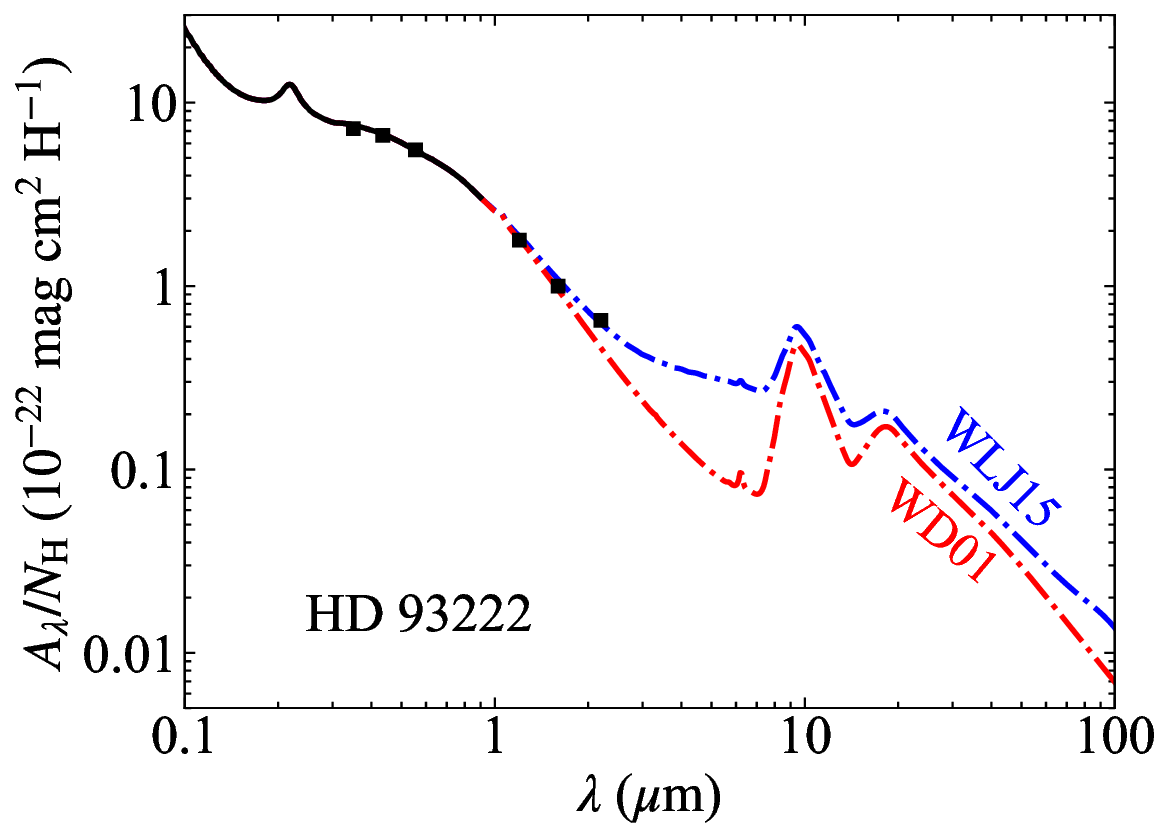}
\end{center}
\vspace{-4mm}
\caption{\label{fig:extcurv}\footnotesize
         The interstellar extinction curve
         from the far-UV to the far-IR
         for the line of sight toward HD\,93222,
         with the FM88 parameterization
         for $\lambda^{-1}>3.3\mum^{-1}$,
         the CCM parameterization (with $R_V=4.76$)
         for  $1.1\mum^{-1} < \lambda^{-1} < 3.3\mum^{-1}$,
         and the $R_V=3.1$ model curves of
         Weingartner \& Draine (2001; red dashed line)
         and Wang, Li \& Jiang (2015; blue dot-dashed line)
         for $\lambda > 0.9\mum$.
         The U, B, V, J, H, K photometric extinction
         data points are superimposed on the extinction
         curve as black squares.
         }
\vspace{-3mm}
\end{figure*}

For $\lambda^{-1} < 1.1\mum^{-1}$, we approximate
the extinction curve by the model extinction calculated
from the standard silicate-graphite-PAH model of
Weingartner \& Draine (2001; WD01) for $R_V=3.1$, or
the extinction calculated by Wang, Li \& Jiang (2015; WLJ15).
The WLJ15 model is essentially an extension of WD01,
but includes an additional population of
very large, micron-sized graphite grains
which are introduced to account for the observed
flat mid-IR extinction at $3 < \lambda < 8\mum$.
Figure~\ref{fig:extcurv} presents the extinction curve
derived for HD\,93222, spanning from the far-UV to
the near-IR. The black squares indicate the broadband
photometric extinction data measured in the U, B, V,
J, H, and K bands. It appears that the WLJ15 extinction
model is a better choice since it better matches the K
band extinction (see Figure~\ref{fig:extcurv}).

\section{Discussion}\label{sec:discussion}
\subsection{The Exceedingly Narrow 2175$\Angstrom$
                    Extinction Bump of HD\,93222}
%
Figure~\ref{fig:extcurv} compares
the extinction curve derived in \S\ref{sec:extcurv}
for HD\,93222 and the CCM representation for
$R_V=4.76$, an optical total-to-selective extinction
ratio of HD\,93222 (Valencic et al.\ 2004).
It is apparent that, with a FWHM
of $\gamma\approx0.76\mum^{-1}$,
the 2175$\Angstrom$ extinction bump
of HD\,93222 is appreciably narrower
and weaker than that of the CCM $R_V=4.76$ curve.
However, the far-UV extinction rise of HD\,93222 is
considerably steeper than that of
the CCM $R_V=4.76$ curve.
As a matter of fact, the far-UV extinction rise
of HD\,93222 resembles that of
the CCM $R_V=3.1$ curve, which is commonly
taken to represent the mean extinction curve
of the Galactic diffuse ISM.
Indeed, the curvature of the HD\,93222 curve
at $\lambda^{-1}>5.9\mum^{-1}$ is essentially
the same as that of the CCM $R_V=3.1$ curve.
Figure~\ref{fig:extcurv} also
shows that the 2175$\Angstrom$ bump
is considerably narrower than that of
the CCM curves of $R_V=4.76$ and $R_V=3.1$.

\begin{figure*}[htp]
\vspace{-1mm}
\begin{center}
\includegraphics[width=9.0cm,angle=0]{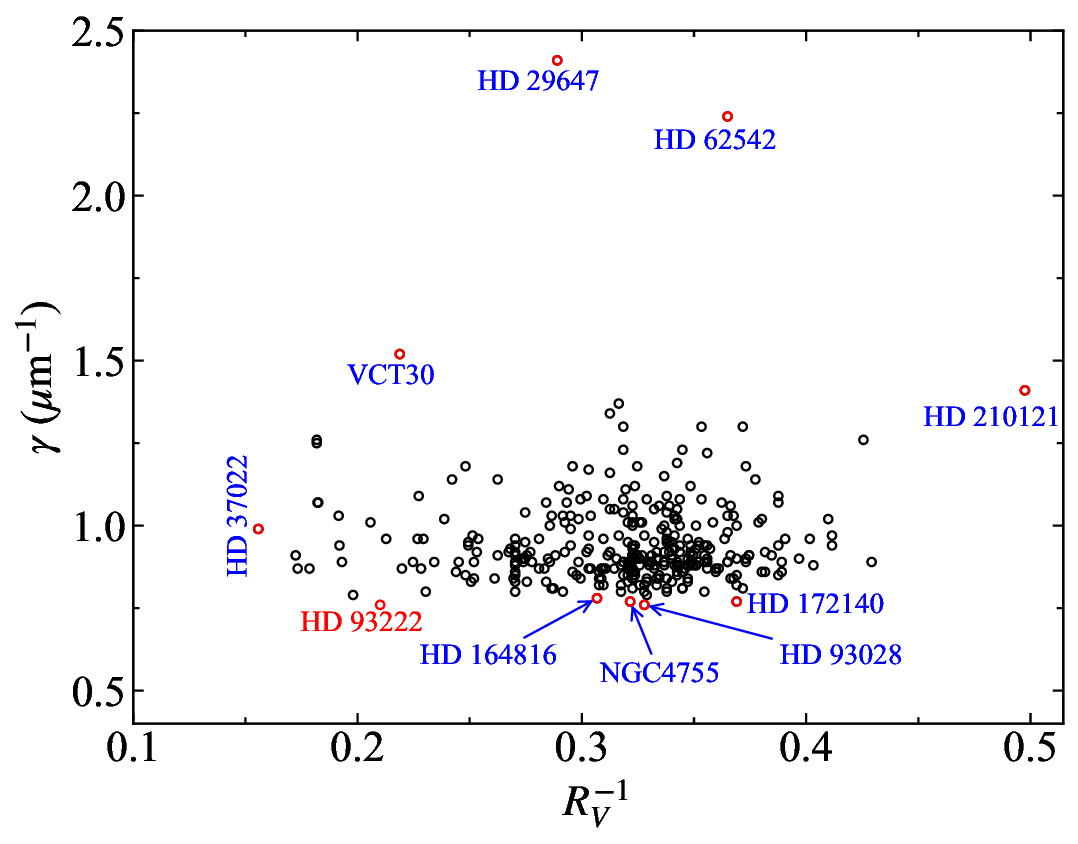}
\end{center}
\vspace{-4mm}
\caption{\label{fig:FM07}
The 2175$\Angstrom$ extinction bump width
(FWHM; $\gamma$) plotted against $R_V^{-1}$
for the 328 Galactic interstellar extinction curves of
Valencic et al.\ (2004) and Fitzpatrick \& Massa (2007).
The labelled sightlines (red circles) show those
exhibiting extreme $\gamma$ and $R_V$ values.
         }
\vspace{-3mm}
\end{figure*}

HD\,93222 serves as a striking example of
sightlines exhibiting an unusually narrow
extinction bump at 2175$\Angstrom$.
The extinction curve of HD\,93222
derived in \S\ref{sec:extcurv} shows a bump
centered at $\simali$2190$\Angstrom$,
which only slightly deviates from the canonical
2175$\Angstrom$. The most pronounced
characteristics of the extinction bump of
HD\,93222 is its width:
with $\gamma\approx0.76\mum^{-1}$,
it is considerably narrower than that of
the Galactic diffuse ISM for which the average
width is $\simali$0.92$\mum$
(Valencic et al.\ 2004, Fitzpatrick \& Massa 2007).
To our knowledge, the extinction bump of HD\,93222
is one of the narrowest ever observed in the Galactic
interstellar lines of sight. Figure~\ref{fig:FM07} illustrates
those five sightlines of which the extinction bump is
among the narrowest in 328 Galactic interstellar sources
(Valencic et al.\ 2004):
HD\,164816 ($\gamma\approx0.78\mum$),
HD\,172140 ($\gamma\approx0.77\mum$),
NGC\,4755 ($\gamma\approx0.77\mum$),
HD\,93028 ($\gamma\approx0.76\mum$), and
HD\,93222 ($\gamma\approx0.76\mum$).
Figure~\ref{fig:FM07} also compares
the bump width $\gamma$ with $R_V^{-1}$.
It is clear that $\gamma$ exhibits no obvious
relationship with $R_V^{-1}$, consistent with
the earlier finding of Cardelli et al.\ (1989).


The width of the 2175$\Angstrom$ bump is thought
to reflect the size distribution, structural order,
and processing history of the bump carrier
(and other dust components as well).
The unusual characteristics of the extinction bump
observed toward HD\,93222 may indicate atypical
physical conditions along its line of sight,
including strong processing by UV starlight
from star formation activities,
selective fragmentation of grain structures,
or a deficiency in large aromatic carbonaceous compounds.
This exceptionally narrow bump thus provides
a valuable test case for constraining dust models
and offers unique insights into the physical nature
and origin of the 2175$\Angstrom$ extinction bump.

\subsection{Modeling the Extinction Curve}\label{sec:extmod}
%
%
%
%
%
We fit the extinction curve of HD\,93222
in terms of the standard silicate–graphite
interstellar grain model. For simplicity,
we assume the dust to be spherical in shape.
We adopt an exponentially-cutoff power-law size
distribution for both components:
$dn_i/da = \nH\,B_i\,a^{-\alpha_i} \exp\left(-a/a_{c,i}\right)$
for the size range of
$50\Angstrom < a < 2.5\mum$,
where $a$ is the spherical radius of the dust,
$\nH$ is the number density of H nuclei,
$dn_i$ is the number density of dust of type $i$
with radii in the interval [$a$, $a$\,$+$\,$da$],
$\alpha_i$ and $a_{c,i}$ are respectively
the power index and exponential cutoff size
for dust of type $i$, and
$B_i$ is the constant related to
the total amount of dust of type $i$.
The total extinction per H column
at wavelength $\lambda$ is given by
\begin{equation}\label{eq:Amod}
A_\lambda/\NH = 1.086
            \sum_i \int da \frac{1}{\nH}
            \frac{dn_i}{da}
            C_{{\rm ext},i}(a,\lambda),
\end{equation}
where the summation is over the two grain types
(i.e., amorphous silicate and graphite),
and $C_{{\rm ext},i}(a,\lambda)$
is the extinction cross section of
grain type $i$ of size $a$
at wavelength $\lambda$
which can be calculated
from Mie theory (Bohren \& Huffman 1983)
using the dielectric functions of
``astronomical'' silicate and graphite
of Draine \& Lee (1984).

\begin{figure*}[htp]
\vspace{-1mm}
\begin{center}
\includegraphics[width=8cm,angle=0]{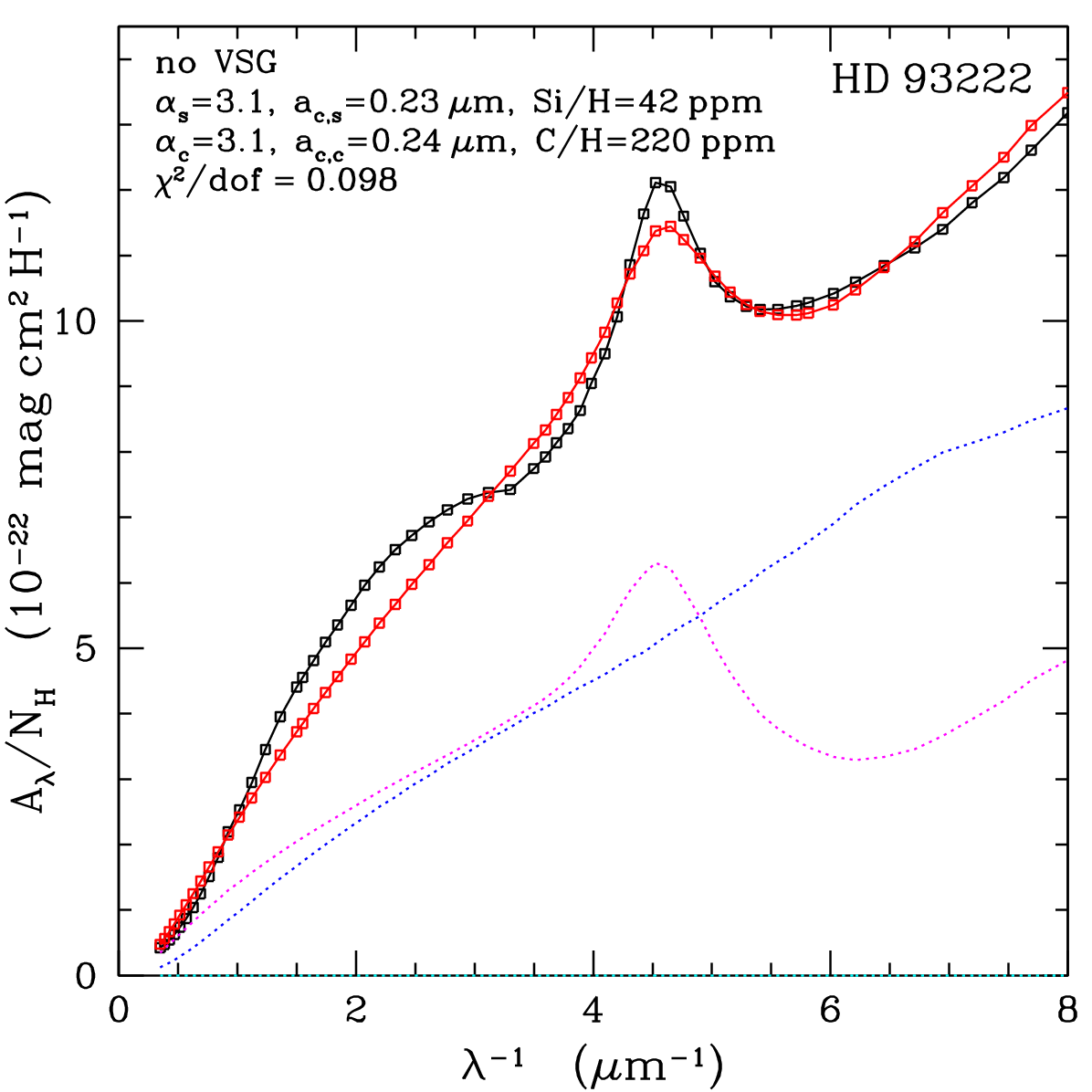}
\includegraphics[width=8cm,angle=0]{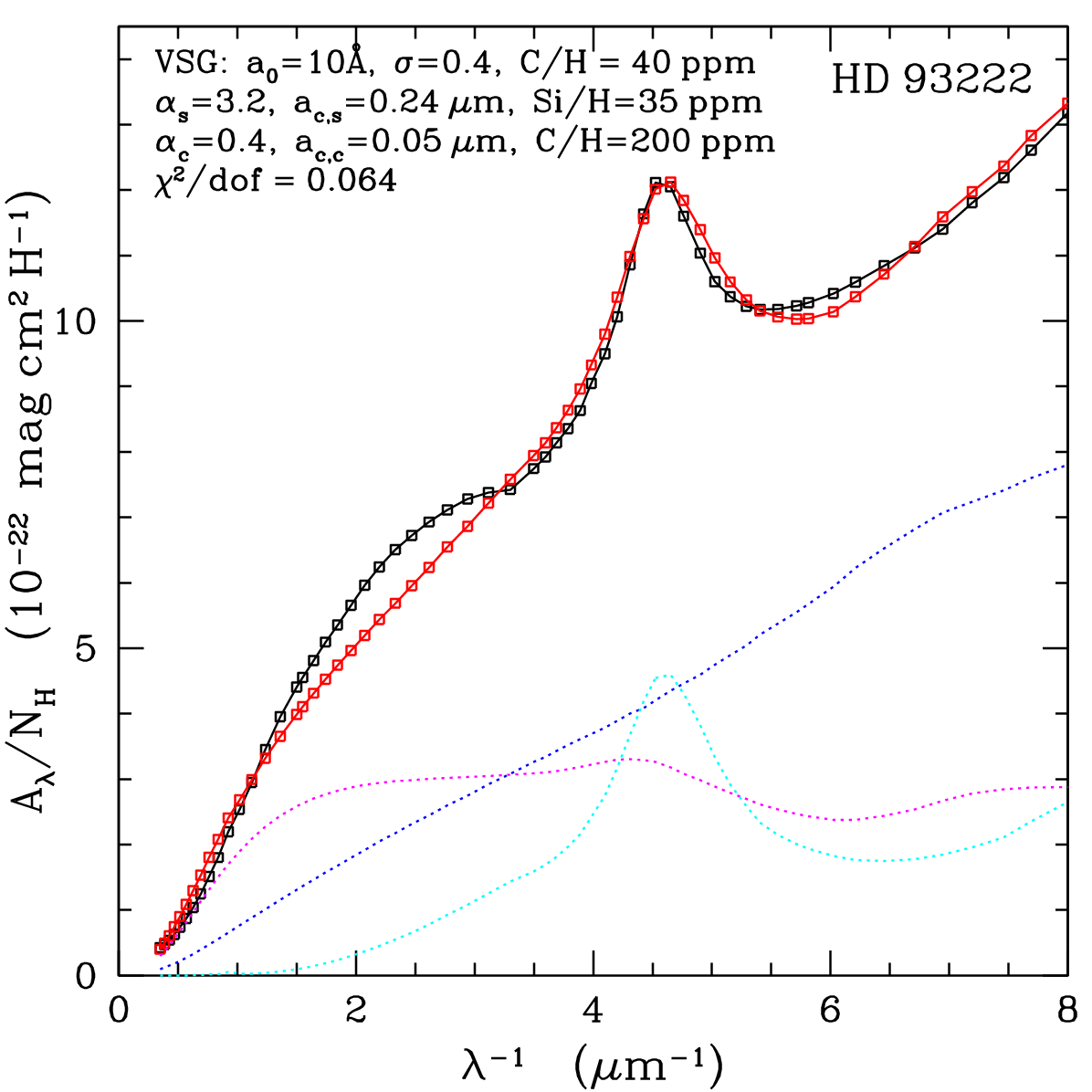}
\end{center}
\vspace{-4mm}
\caption{\label{fig:dustmodel1}
Left panel (a): Fitting the extinction curve of
HD\,93222 (black squares) with a mixture
of silicate (blue dashed line) and graphite grains
(magenta dashed line), each with an
exponentially-cutoff power-law size distribution
(red squares). Right panel (b):  Same as (a) but with
an additional population of nano graphite grains
(cyan dashed line) which lock up a carbon abundance
(relative to H) of $\CTOH=40\ppm$.
``VSG'' refers to very small grains
(i.e., nano graphite grains).
The right panel represents our preferred model.
         }
\vspace{-3mm}
\end{figure*}

Figure~\ref{fig:dustmodel1}a compares
the best-fit model extinction curve with
that observationally derived for HD\,93222.
The best-fit model parameters
(i.e., Model \#1) are tabulated
in Table~\ref{tab:model}.
It is apparent that, although the overall fit
is not too bad, the model predicts an extinction
bump appreciably too broad to agree with
the observed one. Also, the model considerably
deviates from the observed one
at $1.1 < \lambda^{-1} < 4.2 \mum^{-1}$.
It is worth noting that the difficulty of fitting
the extinction curve of HD\,93222 derived by
Valencic et al.\ (2004) has already been
recognized by Zuo et al.\ (2021b).

To reproduce the observed narrow 2175$\Angstrom$
extinction bump, we introduce an additional dust
component: a population of nano-sized graphite
grains (or ``very small grains'', VSG)
with a log-normal size distribution,
$dn/d\ln a \propto \exp\{-[{\rm ln}(a/a_0)]^2/(2\sigma^2)\}$,
where $a_0$ and $\sigma$ characterize the peak
and width of this distribution.\footnote{%
  For this distribution, $a^3 dn/d\ln a$ peaks
  at $a_p = a_0\,\exp\left(3\sigma^2\right)$.
  }
Let $\CTOH$ be the amount of C (relative to H)
tied up in the nano graphitic grain component.
Figure~\ref{fig:dustmodel1}b shows the best-fit
model which includes, in addition to amorphous
silicate and graphite grains, an extra population
of nano graphitic grains of $\CTOH=40\ppm$.
We take $a_0=10\Angstrom$ and $\sigma=0.4$.
The exact choice of $a_0$ and $\sigma$ does not
matter much as long as this extra component is
nano-sized so that the nano-graphitic grains are
in the Rayleigh regime
(i.e., $2\pi\,a/\lambda\ll 1$).\footnote{%
  In the Rayleigh regime, the extinction is
  dominated by absorption and the absorption
  cross section (on a per unit volume basis) is
  independent of grain size.
  }

\begin{figure*}[htp]
\vspace{-1mm}
\begin{center}
  \includegraphics[width=8cm,angle=0]{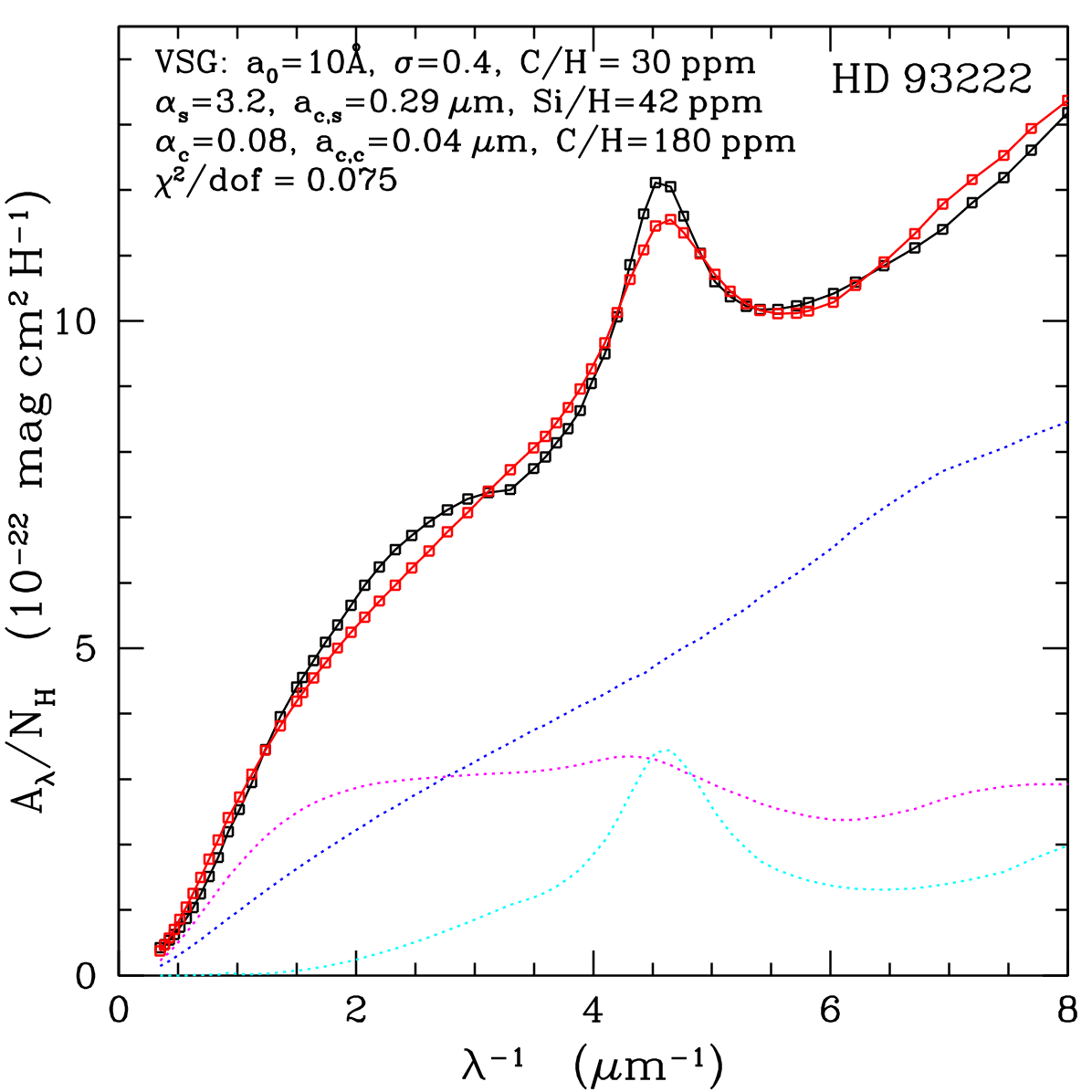}
   \includegraphics[width=8cm,angle=0]{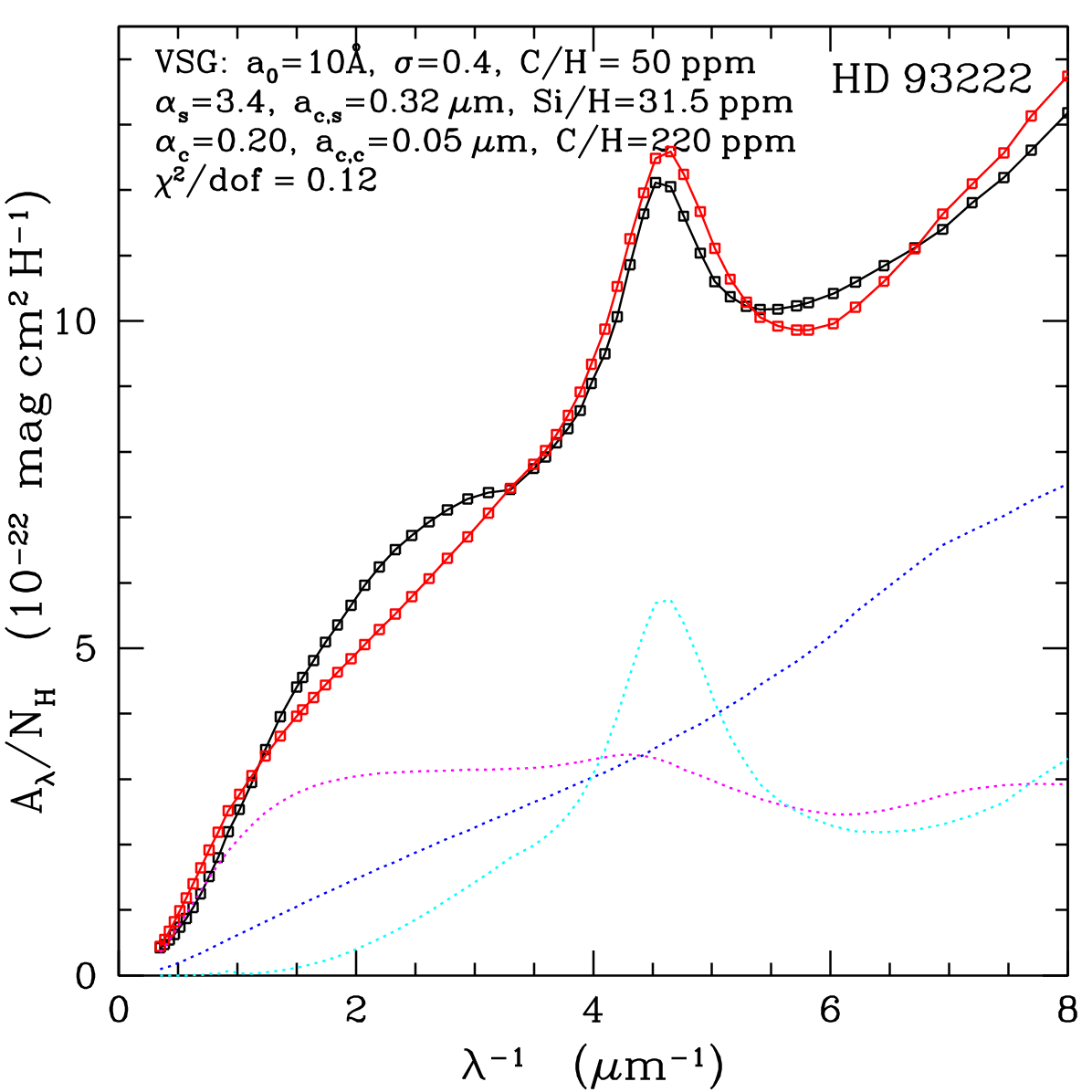}
\end{center}
\vspace{-4mm}
\caption{\label{fig:dustmodel2}
Left panel (a): Same as Figure~\ref{fig:dustmodel1}b
but with  $\CTOH=30\ppm$.
Right panel (b): Same as Figure~\ref{fig:dustmodel1}b
but with  $\CTOH=50\ppm$.
         }
\vspace{-3mm}
\end{figure*}

As shown in Figure~\ref{fig:dustmodel1}b,
the inclusion of a population of nano graphite grains
significantly improves the fit to the observed
extinction curve, including the narrow
2175$\Angstrom$ extinction bump.
Table~\ref{tab:model} also tabulates
the corresponding model parameters
(see Model \#2).
We admit that the model
extinction curve deviates from the ``observed''
one at $1.2 < \lambda^{-1} < 3.3 \mum^{-1}$.
We note that the ``observed'' extinction curve
in this range is not really derived from observations.
It was constructed from the $R_V$-based CCM
parameterization (see \S\ref{sec:extcurv}).
As the CCM parameterization may not be valid
for every individual sightline, we shall not worry
too much about the deviation at this range.
Finally, we have also explored different choices
of $\CTOH$. Figure~\ref{fig:dustmodel2} shows
the best-fit extinction model curves obtained
with $\CTOH=30\ppm$
(see Figure~\ref{fig:dustmodel2}a)
and 50$\ppm$ (see Figure~\ref{fig:dustmodel2}b).
Apparently, the $\CTOH=30\ppm$ model
under-predicts the observed 2175$\Angstrom$
extinction bump, while  the $\CTOH=50\ppm$ model
over-predicts the observed bump.
The model parameters are also tabulated
in Table~\ref{tab:model}
(see Models \#3 and \#4).

We note that, in principle, the interstellar extinction
curve of HD\,93222 could also be modeled
in terms of composite dust consisting of vacuum
and small silicate and amorphous carbon grains
(Mathis 1996), or silicate core-carbon mantle grains
(Li \& Greenberg 1997, Jones et al.\ 2013).
As these models all attribute the 2175$\Angstrom$
bump to small graphitic grains, we expect
the quantities and size distributions of
these small graphitic grains to be essentially
the same as that derived here.
The ``astrodust'' model of Hensley \& Draine (2023)
assumes an ``astrodust'' component
which resembles the composite dust of Mathis (1996),
combined with an additional population of
``astronomical'' polycyclic aromatic hydrocarbon
(PAH) molecules to account for the 2175$\Angstrom$ bump.
As the optical properties of astro-PAHs
were ``designed'' to produce an extinction bump
with a fixed width of 1.0$\mum^{-1}$
(see Li \& Draine 2001), the astrodust+PAH model
is not expected to reproduce the narrow bump
seen in HD\,93222. However, this does not mean
that PAHs are not able to explain the HD\,93222
extinction bump. Totally in the opposite, it is actually
very likely that mixtures of specific PAH molecules
of different sizes, structures and charging states
are expected to be capable of reproducing the bump
detected in HD\,93222,
as demonstrated by Lin et al.\ (2023, 2025)
for the Galactic diffuse ISM and JADES-GS-z6-0,
a distant galaxy at $z\approx6.71$.

\subsection{The Interstellar Environments around HD\,93222}
\label{sec:environment}
\begin{figure*}[htp]
\vspace{-1mm}
\begin{center}
\includegraphics[width=10.0cm,angle=0]{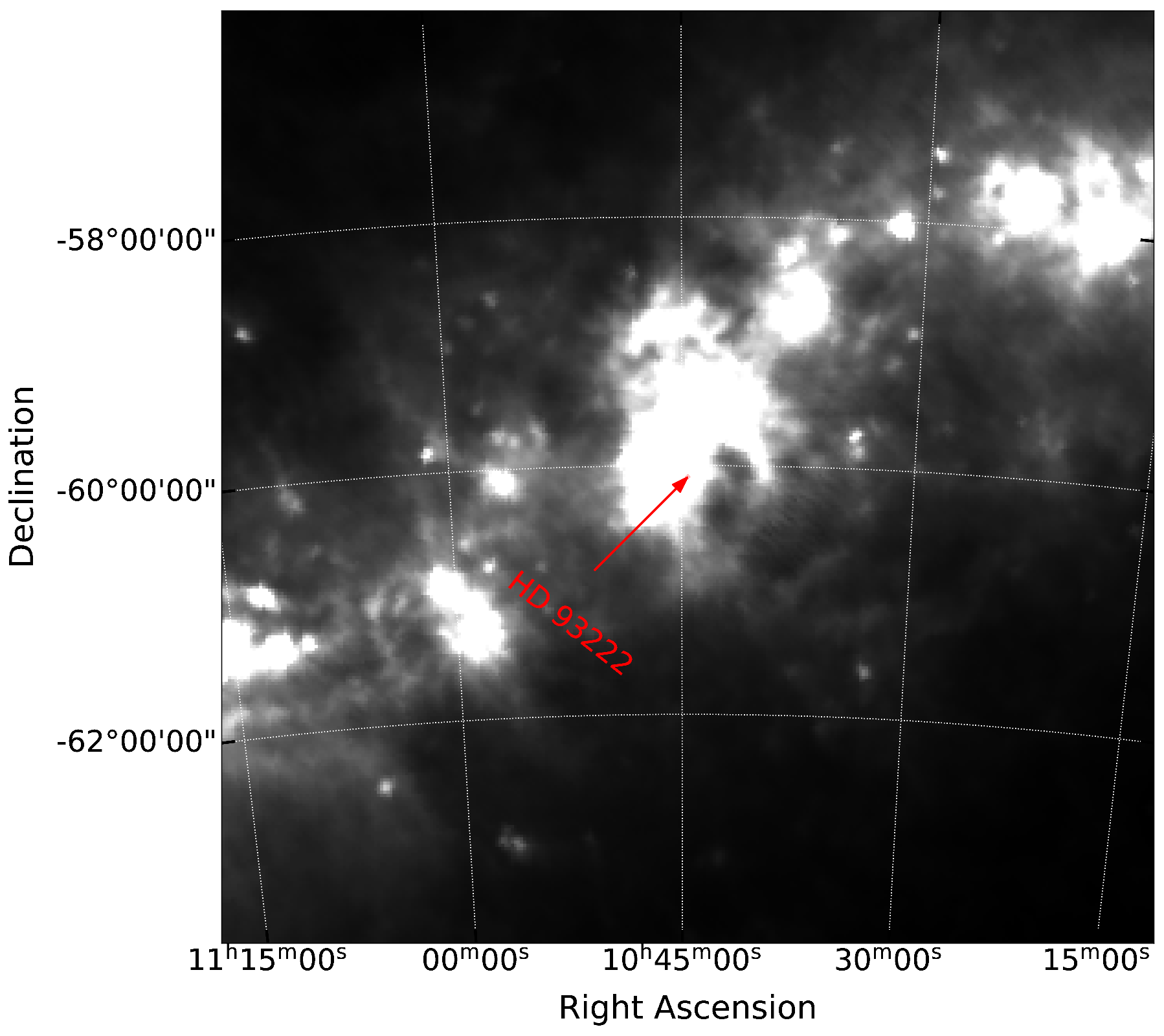}
\end{center}
\vspace{-4mm}
\caption{\label{fig:iras100map}
  IRAS 100$\mum$ map of the Carina nebula.
  The position of HD\,93222 is also shown.
         }
\vspace{-3mm}
\end{figure*}

As illustrated in Figure~\ref{fig:iras100map},
a dust thermal emission map of the Carina nebula
obtained by the {\it Infrared Astronomical Satellite}
(IRAS) at 100$\mum$,
HD~93222 is located in the southern part of
the base of the Carina nebula,
one of the largest star-forming regions
in the Milky Way, which hosts numerous
bright O-type stars. Walborn et al.\ (1982)
detected a strong Mg\,\textsc{ii}
and C\,\textsc{ii} absorption component
at $-350\km\s^{-1}$ in the IUE data,
suggesting a connection with diffuse X-ray emission
and near-UV nebulosity.
Compared to the other stars in the region,
HD\,93222 exhibits more discrete velocity structures
in its high-ionization lines, indicating a more turbulent
environment and stronger stellar winds.
The strong radiation field and intense stellar winds
around HD\,93222 likely lead to the fragmentation
of large dust grains into smaller ones and eventually
the production of nano graphite grains
and a pronounced, sharp 2175$\Angstrom$
extinction bump.

Also, the prominent C\,\textsc{ii} feature
suggests that the surrounding medium
is relatively carbon-rich, facilitating the formation
of small-sized carbonaceous dust grains
widely believed to be responsible for
the 2175$\Angstrom$ extinction bump.
The narrow and intense bump observed
toward HD\,93222 supports this scenario.
Moreover, the steep rise in the far-UV extinction
(compared to that predicted from the CCM
parametrization with $R_V=4.76$) is also
attributed to the abundance of small grains,
which are more efficiently produced under such
harsh conditions.
These factors together contribute to the unique
extinction curve observed along the line of sight
to HD\,93222.

We have so far adopted $R_V=4.76$ which was
derived by Valencic et al.\ (2004) from the JHK colors
using the empirical relations of Fitzpatrick (1999).
Although the $R_V$ value derived from such an
approach may be subject to large uncertainty,
Fitzpatrick \& Massa (2007) analyzed the spectral
energy distribution of HD\,93222 and derived
$R_V\approx5.05$, close to that of Valencic et al.\ (2004).
In general, the value of $R_V$ depends upon
the environment along the line of sight.
A direction through low-density ISM
usually has a rather low $R_V$ ($\simali$3.1).
Lines of sight penetrating into a dense cloud,
such as the Ophiuchus or Taurus molecular clouds,
usually show $4 < R_V < 6$. However, it is difficult
to estimate $R_V$ quantitatively from
the environment of a line of sight.
We note that Cygni OB2 No.\,12 lies behind
a dense cloud of dust but has $R_V\approx3.1$;
also, parts of the Taurus molecular cloud
have $R_V$\,$\simali$3.0--3.5 as well
(see Mathis 1990 and references therein).
Valencic et al.\ (2004) also cautioned that
$R_V$ should not be assigned a physical meaning.
Indeed, a large value of $R_V\approx4.76$
is more appropriate for molecular clouds,
the relatively small reddening of
$E(B-V)\approx0.35\magni$
for the HD\,93222 line of sight
(Fitzpatrick et al.\ 2019)
and negligible molecular content\footnote{%
  With an atomic H column density of
  $N({\rm HI})\approx2.95\times10^{21}\cm^{-2}$
  and an H$_2$ column density of
  $N({\rm H}_2)\approx0.059\times10^{21}\cm^{-2}$
  (Jenkins 2019), the molecular fraction of
  the HD\,93222 sightline is only
  $f({\rm H}_2)=2N({\rm H}_2)/
  \left[2N({\rm H}_2)+N({\rm HI})\right]\approx3.8\%$.
  With $E(B-V)=0.35\magni$,
  the extinction-to-gas ratio for the HD\,93222 sightline is
  $A_V/\NH\approx5.44\times10^{-22}\magni\cm^2\,\rmH^{-1}$
  for $R_V=4.76$,
  and $\approx3.65\times10^{-22}\magni\cm^2\,\rmH^{-1}$
  for $R_V=3.2$, both of which are within the typical
  interstellar ranges of extinction-to-gas ratios
  (e.g., see Zuo et al.\ 2021a).
  }
suggest that the sightline toward HD\,93222 likely
samples diffuse gas.

Alternatively, the actual $R_V$ value for
the HD\,93222 sightline may be appreciably
lower than 4.76.
Indeed, the $R_V$ value derived from the near-IR
photometry tends to be high
(e.g., see Patriarchi et al.\ 2001),
whereas, in sharp contrast, $R_V$ based on direct
high quality optical photometry yields
$R_V\approx3.19$ (Neckel et al.\ 1980,
Jensen \& Snow 2007).
Nevertheless, as shown in
eqs.\,\ref{eq:fit}--\ref{eq:FarUV},
$R_V$ is not involved in the FM88
parametrization;
i.e., for $\lambda^{-1} > 3.3\mum^{-1}$,
the shape of the extinction curve
and the profile of the 2175$\Angstrom$
bump are independent of the choice of $R_V$.

\begin{figure*}[htp]
\vspace{-1mm}
\begin{center}
\includegraphics[width=12cm,angle=0]{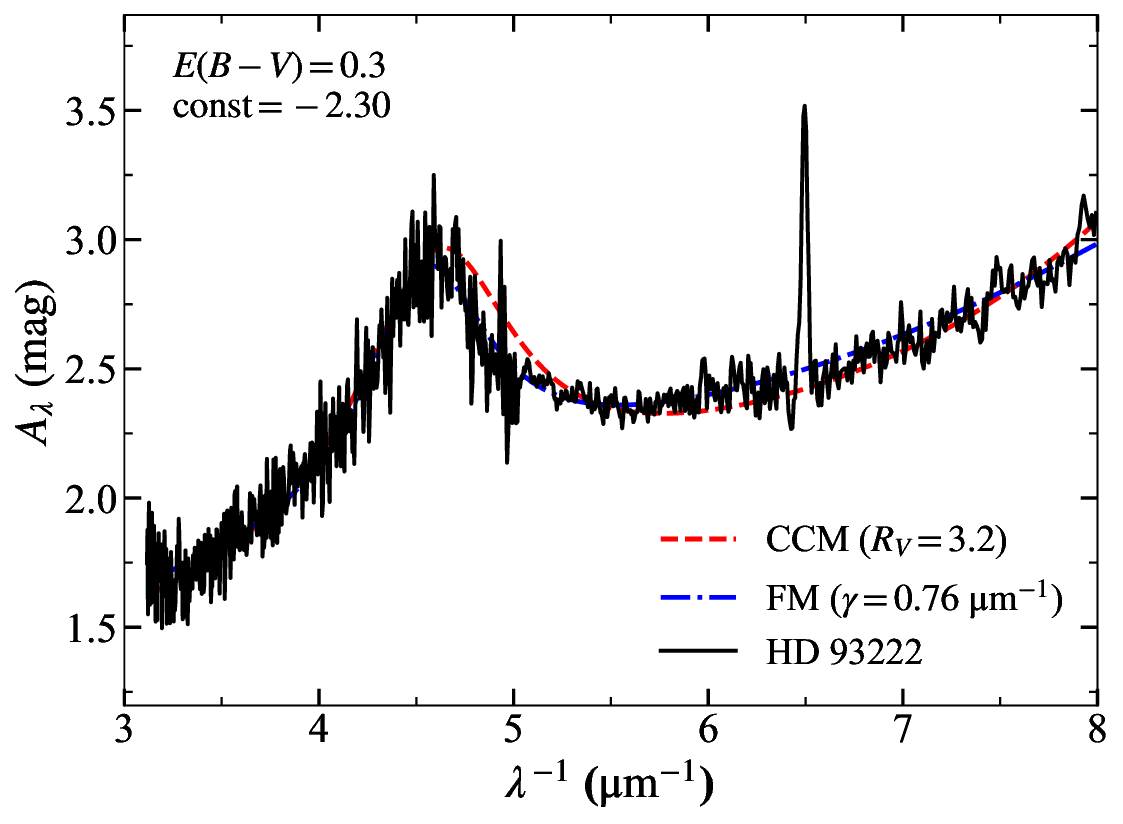}
\end{center}
\vspace{-4mm}
\caption{\label{fig:pair}\footnotesize
The extinction curve derived from the traditional
``pair method'' (solid black line)
by comparing the IUE spectrum of HD\,93222
with that of the comparison star HD\,47839,
which differ in $E(B-V)$ by $\simali$0.3$\magni$
(see \S\ref{sec:environment}).
Also shown are the CCM $R_V=3.2$ curve
(red dashed line) and the FM88 curve
(blue dot-dashed line).
With a vertical shift of $\constant=-2.30$
(see eq.\,\ref{eq:pair}), the extinction curve
derived from the ``pair method'' matches
the CCM $R_V=3.2$ curve very well
(except the latter has a broader 2175$\Angstrom$ bump).
In contrast, the FM88 curve with a narrow
bump of $\gamma=0.76\mum^{-1}$
closely fits the ``pair-method''-derived
extinction curve.
           }
\vspace{-3mm}
\end{figure*}

\begin{figure*}[htp]
\vspace{-1mm}
\begin{center}
\includegraphics[width=12cm,angle=0]{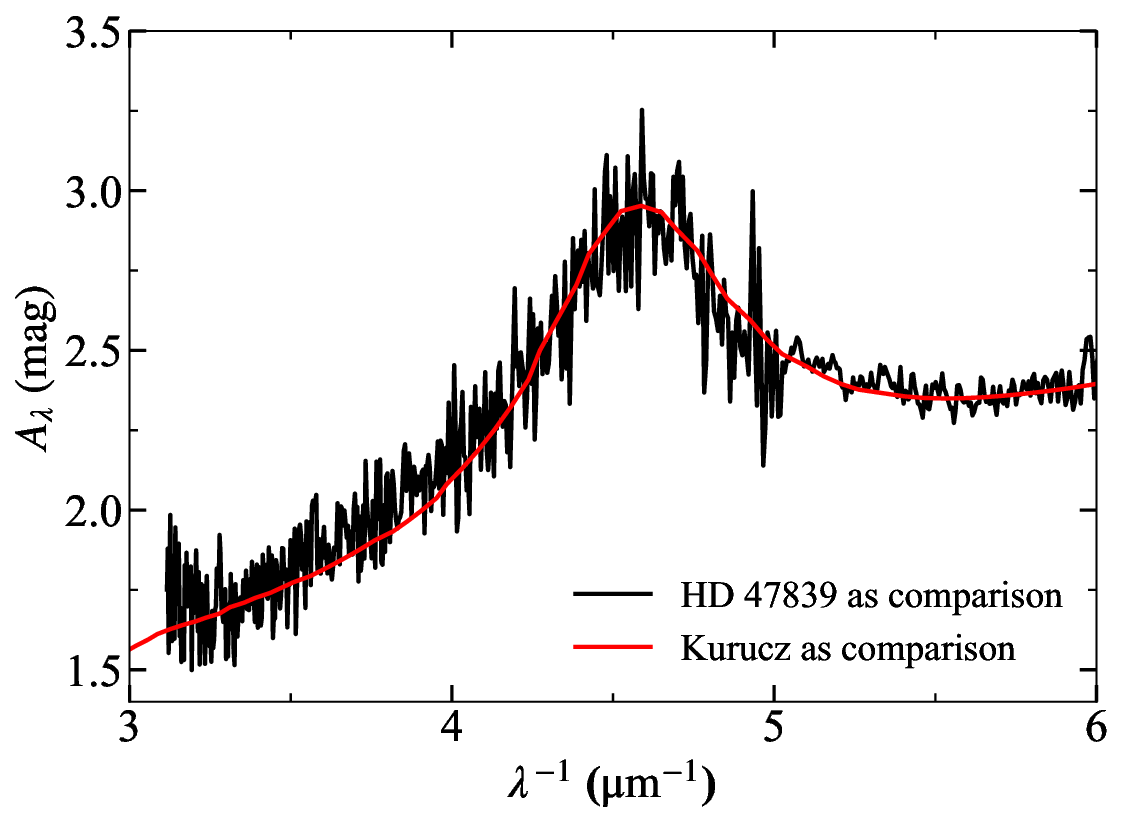}
\end{center}
\vspace{-4mm}
\caption{\label{fig:pair_kurucz}\footnotesize
Comparison of the extinction curve
derived from the ``pair method''
(solid black line; see \S\ref{sec:environment})
with that derived based on the Kurucz model
spectrum (solid red line; see \S\ref{sec:extcurv}),
after vertically-shifted.
           }
\vspace{-3mm}
\end{figure*}

It is interesting to note that, as will be shown
below, the HD\,93222 extinction curve derived
from the traditional ``pair method'' is actually
well represented by the CCM $R_V=3.2$ parametrization.
The ``pair method'' involves photometric or spectrophotometric
observations of two stars of identical spectral types,
with one star located behind a dust cloud and
another star, (in ideal case) unaffected by interstellar dust,
so that there is no obscuration between the observer and the star.
Let $\Ftarget$ be the observed flux from
the reddened, target star at a distance of $\dtarget$,
and $\Fcomp$ be the flux from the unreddened,
comparison star at a distance of $\dcomp$.
If both stars are of identical spectral and luminosity types,
the extinction $A_\lambda$---measured in
``magnitudes''--- is
\begin{equation}\label{eq:pair}
A_\lambda = 2.5
\log\left\{\frac{\Fcomp}{\Ftarget}\right\}
+ 5 \log\left\{\frac{\dcomp}{\dtarget}\right\}~~.
\end{equation}
In our case, the reddened, target star is HD\,93222.
We take HD\,47839 (also known as 15 Mon or S Mon),
a brilliant, massive O-type star located in the constellation
Monoceros, as the comparison star.
If we know accurately the distances of both the target
and comparison stars, we can derive from eq.\,\ref{eq:pair}
the extinction curve of HD\,93222
at $\lambda^{-1}>3.3\mum^{-1}$
by comparing the IUE spectrum of HD\,93222
with that of HD\,47839.
In Figure~\ref{fig:pair} we show
$A_\lambda = 2.5\log\left\{\Fcomp/\Ftarget\right\}\,+\,\constant$,
the extinction curve of HD\,93222
derived from the ``pair method'',
with a constant offset ($\constant$)
to account for the often unknown distance term
$5\log\left\{\dcomp/\dtarget\right\}$.
The constant offset ($\constant$) is determined
by matching $A_\lambda$ with $\Accm$,
the extinction predicted from the $R_V$-based
CCM parametrization for HD\,93222,
where $\Accm = E(B-V)\times
\left\{a(x)\,R_V+b(x)\right\}$,
$E(B-V)\approx0.30\magni$ is
the {\it reddening difference}
between HD\,93222 and HD\,47839,
and the wavelength-dependent coefficients
$a(x)$ and $b(x)$ are given in CCM.
As shown in Figure~\ref{fig:pair},
with $R_V=3.2$, we are able to closely match
$A_\lambda$ and $\Accm$,
except the $\Accm$ curve predicts a broader
2175$\Angstrom$ extinction bump.
We have also applied the FM88 parameterization
to fit the HD\,93222 extinction curve derived
from the ``pair method''. Figure~\ref{fig:pair} clearly
shows that a width of $0.76\mum^{-1}$
as derived earlier (see \S\ref{sec:extcurv})
describes well the ``pair-method''-derived
2175$\Angstrom$ bump.
Again, this demonstrates that the extreme
narrowness of the 2175$\Angstrom$ bump
derived for HD\,93222 is not affected by $R_V$.
Finally, we compare the extinction curve derived
from the ``pair method'' with that derived based
on the Kurucz model spectrum (see \S\ref{sec:extcurv}).
As shown in Figure~\ref{fig:pair_kurucz},
the vertically-shifted extinction curves derived from these two
methods, particularly the derived 2175$\Angstrom$
extinction bumps, closely resemble each other.
This demonstrates the robustness of the extreme
narrowness of the 2175$\Angstrom$ bump
of the HD\,93222 sightline.

\section{Summary}\label{sec:summary}
We have determined the extinction curve
along the line of sight toward HD\,93222
in the Carina nebula which shows an exceedingly
narrow 2175$\Angstrom$ extinction bump.
With a width of $\gamma\approx0.76\mum^{-1}$,
the extinction bump of the HD\,93222 sightline
is among the narrowest ever observed in the Milky Way.
The derived extinction curve is modeled in terms of
the standard silicate-graphite interstellar grain model.
It is found that, to explain the observed extinction
characteristics of HD\,93222, in addition to the conventional
silicate and graphite dust mixture, an extra component of
nano-sized graphitic grains is required.
We argue that this is likely related to the strong radiation field
and intense stellar winds around HD\,93222.

\acknowledgments
We thank the anonymous referee for his/her
helpful and stimulating comments and suggestions
which considerably improved the quality
and presentation of this paper.
QW and XJY are supported in part by NSFC\,12333005
and 12122302, CMS-CSST-2021-A09,
Hebei NSF\,A2023205036,
the Innovative Research Group Project of
Natural Science Foundation of Hunan Province of China No. 2024JJ1008,
and a Postgraduate Scientific Research Innovation Project
of Xiangtan University (No.\,XDCX2023Y155).
This work is based on observations made with the
International Ultraviolet Explorer (IUE).
The data presented in this article were obtained from
the Mikulski Archive for Space Telescopes (MAST) at
the Space Telescope Science Institute.
The specific observations analyzed can be accessed
via \dataset[doi:DOI]{https://doi.org/10.17909/ha4q-2y11}.


%

\begin{table*}[htp]
\centering
\captionsetup{
  singlelinecheck=off,
  justification=centering,
  skip=6pt,  
  font=normalsize  
}
\caption{Stellar Parameters of HD\,93222}
\label{tab:stellarpara}
\normalsize
\setlength{\tabcolsep}{8pt}  
\renewcommand{\arraystretch}{1.3}  
\begin{tabular}{lcr}
\toprule
Spectral Type & O7 IIIf & Fitzpatrick et al.\ (2019) \\
$E(B-V)$ (mag) & 0.35 & Fitzpatrick et al.\ (2019)\\
$A_V$ (mag)     & 1.71 & Fitzpatrick et al.\ (2019) \\
$T_{\mathrm{eff}}$ (K) & 38,000 & Markova et al.\ (2018) \\
$\log g$ (cm$\s^{-2}$)         & 3.9        & Markova et al.\ (2018) \\
$\mathrm{[M/H]}$  & $-0.24$ & Fitzpatrick et al.\ (2019) \\
Distance (pc)          & 2,600      & Markova et al.\ (2018) \\
$\log (L_\bigstar/L_{\odot})$ & 5.36  & Markova et al.\ (2018) \\
$R_\bigstar/R_{\odot}$        & 11     & Markova et al.\ (2018) \\
\bottomrule
\end{tabular}
\vspace{3pt}  
\end{table*}

\begin{table*}[htp]
\centering
\captionsetup{
  singlelinecheck=off,
  justification=centering,
  skip=6pt,  
  font=normalsize  
}
\caption{
Extinction Parameters for HD\,93222
}
\label{tab:Fitpara}
\normalsize
\setlength{\tabcolsep}{8pt}  
\renewcommand{\arraystretch}{1.3}  
\begin{tabular}{lr}
\toprule
$x_0$ ($\mu$m$^{-1}$)     & $4.57 \pm 0.01$ \\
$x_0$ (\AA)               & $2190 \pm 0.01$ \\
$\gamma$ ($\mu$m$^{-1}$)  & $0.76 \pm 0.01$ \\
$c_1$                     & $0.61 \pm 0.01$ \\
$c_2$                     & $0.13 \pm 0.00$ \\
$c_3$                     & $0.31 \pm 0.00$ \\
$c_4$                     & $0.12 \pm 0.01$ \\
$\mathcal{X}^2$           & $3.00$ \\
\bottomrule
\end{tabular}
\end{table*}

\begin{table*}[htp]
\centering
\captionsetup{
  singlelinecheck=off,
  justification=centering,
  skip=6pt,  
  font=normalsize  
}
\caption{Model Parameters for Fitting
the HD\,93222 Extinction Curve
(with Model \#2 Preferred)
}
\label{tab:model}
\normalsize
\setlength{\tabcolsep}{8pt}  
\renewcommand{\arraystretch}{1.3}  
\begin{tabular}{lcccc}
\toprule
Parameters  & Model \#1  & {\bf Model \#2} & Model \#3 & Model \#4\\
\midrule
$A_V/\NH$ ($10^{-22}$\,mag\,cm$^2$\,H$^{-1}$) & 5.53 &{\bf 5.53} & 5.53 & 5.53\\
Silicate: $\alpha_{\rm S}$            & 3.10  & {\bf 3.20} & 3.20 & 3.40\\
Silicate: $a_{c,{\rm S}}$ ($\mu$m)  & 0.23  & {\bf 0.24} & 0.29 & 0.32\\
Graphite: $\alpha_{\rm C}$   & 3.10  & {\bf 0.40} & 0.08 & 0.20\\
Graphite: $a_{c,{\rm C}}$ ($\mu$m)    & 0.24  & {\bf 0.05} & 0.04 & 0.05\\
$\chi^2$/dof            & 0.098 & {\bf 0.064} & 0.075 & 0.12\\
$\mathrm{[C/H]}_{\rm nano}$ (ppm)             & 0     & {\bf 40}   & 30   & 50\\
$\mathrm{[C/H]}_{\rm dust}$ (ppm)             & 220   & {\bf 200}  & 180  & 220\\
$\mathrm{[Si/H]}_{\rm dust}$ (ppm)            & 42    & {\bf 35}   & 42   & 31.5\\
\bottomrule
\end{tabular}
\end{table*}

\begin{figure*}[htp]
\vspace{-1mm}
\begin{center}
\includegraphics[width=10cm,angle=0]{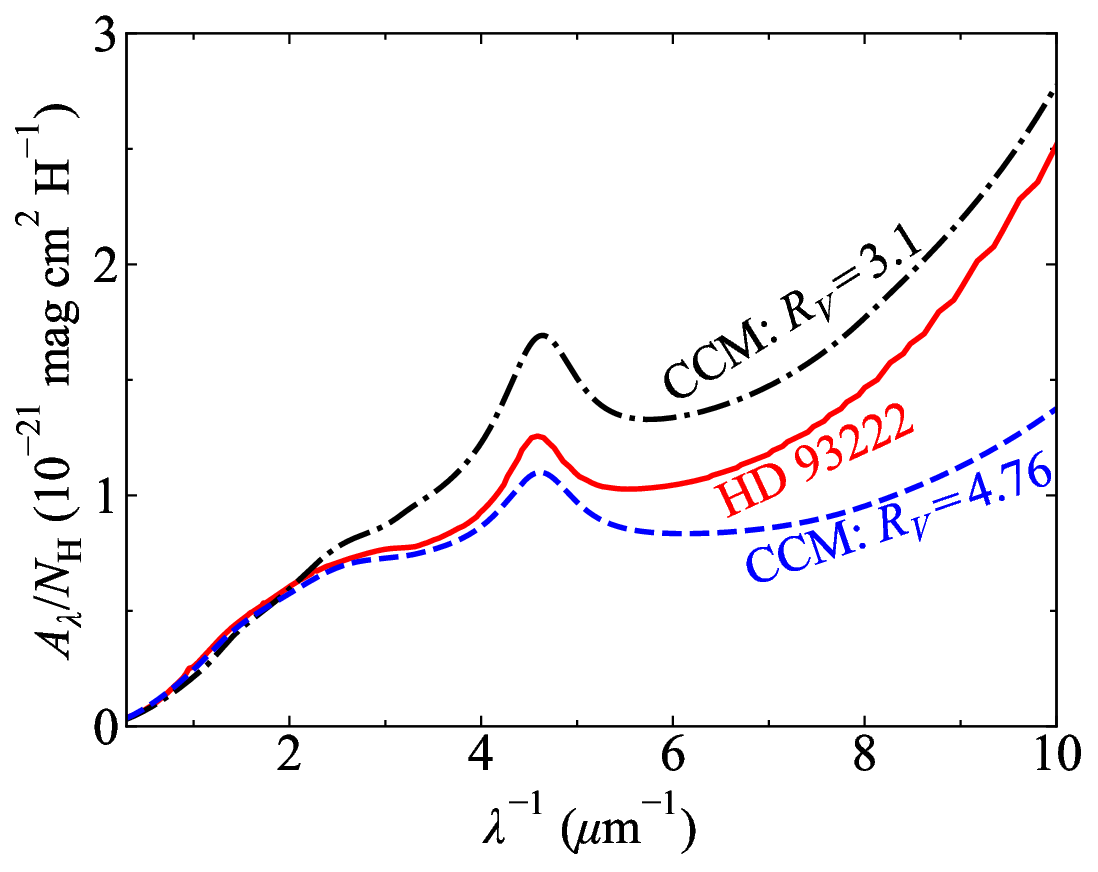}
\end{center}
\vspace{-4mm}
\caption{\label{fig:anomalous extinction}\footnotesize
Comparison of the extinction curve derived
in \S\ref{sec:extcurv} for HD\,93222 (red solid line)
and the CCM representation for $R_V=4.76$
(blue dashed line), the optical total-to-selective
extinction ratio of HD\,93222. Also shown is
the CCM $R_V=3.1$ curve (black dot-dashed line),
which is commonly taken to represent the mean
extinction curve of the Galactic diffuse ISM.
           }
\vspace{-3mm}
\end{figure*}

\clearpage

\end{document}